# Resource Letter RBAI-1: Research-based Assessment Instruments in Physics and Astronomy

Adrian Madsen, Sam McKagan

*American Association of Physics Teachers, College Park, MD*

Eleanor C. Sayre

*Department of Physics, Kansas State University, Manhattan, KS 66506*

**Abstract**

This resource letter provides a guide to research-based assessment instruments (RBAIs) of physics and astronomy content. These are standardized assessments that were rigorously developed and revised using student ideas and interviews, expert input, and statistical analyses. RBAIs have had a major impact on physics and astronomy education reform by providing a universal and convincing measure of student understanding that instructors can use to assess and improve the effectiveness of their teaching. In this resource letter, we present an overview of all content RBAIs in physics and astronomy by topic, research validation, instructional level, format, and themes, to help faculty find the best assessment for their course.

## I. INTRODUCTION

The physics and astronomy education research communities have produced 60+ research-based assessment instruments (RBAIs) of physics and astronomy content, which evaluate the effectiveness of different teaching methods. We define a research-based assessment as an assessment that is developed based on research into student thinking for use by the wider physics and astronomy education community to provide a standardized assessment of teaching and learning. Conceptual RBAIs have had a major impact on physics education reform by providing a universal and convincing measure of student understanding that instructors can use to assess and improve the effectiveness of their teaching. Studies using these instruments consistently show that research-based teaching methods lead to dramatic improvements in students' conceptual understanding of physics.[1,2] These instruments are already being used on a very large scale: The Force Concept Inventory[3] (FCI), a test of basic concepts of forces and acceleration, has been given to thousands of students throughout the world; the use of similar instruments in nearly every subject area of physics is becoming increasingly widespread. According to a recent survey of faculty who are about to participate in the Workshop for New Faculty in Physics and Astronomy, nearly half have heard of the FCI, and nearly a quarter have used it in their classrooms.[4] The use of these instruments has the potential to transform teaching practice by informing instructors about their teaching efficacy so that they can improve it.

Our previous research shows that many physics faculty are aware of the existence of RBAIs for introductory physics, but want to know more about RBAIs for a wider range of topics, including upper-division physics, and about which assessments are available and how to use them.[5] This resource letter addresses these needs of physics faculty by presenting an overview of content RBAIs by topic, research validation, instructional level, format, and themes, to help faculty find the best assessment for their course. A second resource letter will discuss the large number of RBAIs that cover non-content topics such as attitudes and beliefs about physics, epistemologies and expectations, the nature of physics, problem solving, self-efficacy, math skills, reasoning skills and lab skills.

We begin with a general discussion of the process of development and validation of RBAIs (Section II), and then discuss specific RBAIs in each of the major content areas in physics and astronomy. These RBAIs cover a diverse set of topics including mechanics (Section III), electricity and magnetism (Section IV), quantum mechanics and modern physics (Section V), thermodynamics (Section VI), waves and optics (Section VII), and astronomy (Section VIII), at a range of levels from high school to graduate school. The only major physics content area where we are unaware of any RBA is statistical mechanics.

Most RBAIs are multiple-choice tests based on research into students' ideas about a narrow range of introductory-level topics. There are also some assessments of upper-level topics, which are often free-response format, and are based on experts' ideas about a topic, since students' have fewer ideas about these topics coming into the course. Researchers are experimenting with new ways to turn these upper-division free-response RBAIs into multiple-choice assessments. There are also RBAIs which cover a wide range of topics, with fewer questions about each. These can give instructors a better sense of what their students learned about many topics, though, since each topic is not probed in depth, there is more uncertainty in the results.

More details about each of these RBAIs are available at physport.org/assessments, where verified educators can download most RBAIs. Wilcox et al.[6] have a more detailed discussion of upper-division RBAIs.

1.  "Secondary analysis of teaching methods in introductory physics: A 50k - student study," J. Von Korff, B. Archibeque, A. Gomez, S. B. McKagan, E. C. Sayre, E. W. Schenk, C. Shepherd, and L. Sorell, submitted to Am. J. Phys. (2016). (E)

2.  "Interactive-engagement versus traditional methods: A six-thousand-student survey of mechanics test data for introductory physics courses," R.R. Hake, Am. J. Phys. **66** (1), 64–74 (1998). (E)

3.  "Force Concept Inventory," D. Hestenes, M. M. Wells, and G. Swackhamer, Phys. Teach. **30** (3), 141–166 (1992). (E)

4.  "Promoting Instructional Change in New Faculty: An Evaluation of the Physics and Astronomy New Faculty Workshop," C. Henderson, Am. J. Phys. **76** (2), 179–187 (2008). (E)

5.  "Research-based assessment affordances and constraints: Perceptions of physics faculty," A. Madsen, S. B. McKagan, M. S. Martinuk, A. Bell, A. & E. C. Sayre, Phys. Rev. Spec. Top. - Phys. Educ. Res **12** (010115), 1–16 (2015). (E)

6.  "Development and uses of upper-division conceptual assessments," B. R. Wilcox, M. D. Caballero, C. Baily, S. V. Chasteen, Q. X. Ryan, S. J. Pollock, Phys. Rev. Spec. Top. - Phys. Educ. Res **11** (020115), 1–12 (2014). (E)

## II. DEVELOPMENT AND VALIDATION OF RESEARCH-BASED ASSESSMENTS

Good research-based assessment instruments are different from typical exams in that their creation involves extensive research and development by experts in physics education research (PER) and/or astronomy education research (AER) to ensure that the questions represent concepts that faculty think are important, that the possible responses represent real student thinking and make sense to students, and that students' scores reliably tell us something about their understanding. The typical process of developing a research-based assessment includes the following steps:[7,8]

1)      Gathering students' ideas about a given topic, usually with interviews or open-ended written questions.

2)      Using students' ideas to write multiple-choice conceptual questions where the incorrect responses cover the range of students' most common incorrect ideas using the students' actual wording.

3)      Testing these questions with another group of students. Usually, researchers use interviews where students talk about their thinking for each question.

4)      Testing these questions with experts in the discipline to ensure that they agree on the importance of the questions and the correctness of the answers.

5)      Revising questions based on feedback from students and experts.

6)      Administering assessment to large numbers of students. Checking the reproducibility of results across courses and institutions. Checking the distributions of answers. Using various statistical methods to ensure the reliability of the assessment.

7)      Revising again.

This rigorous development process produces valid and reliable assessments that can be used to compare instruction across classes and institutions. Based on the steps to developing a good research-based assessment, we have created list of seven categories of research validation (Table II).

TABLE I. Research validation categories

| |
|---|
| Questions based on research into student thinking |
| Studied with student interviews |
| Studied with expert review |
| Appropriate use of statistical analysis |
| Administered at multiple institutions |
| Research published by someone other than developers |
| At least one peer-reviewed publication |

We determine the level of research validation for an assessment based on how many of the research validation categories apply to the RBA (Table II). RBAIs will have a gold level validation when they have been rigorously developed and recognized by a wider research community. Silver-level RBAIs are well-validated, but only by the developers and not the larger community. This could be because these assessments are new. Bronze-level assessments are those where developers have done some validation but are missing pieces. Finally, research-based validation means that an assessment is likely still in the early stages.

TABLE II. Determination of the level of research validation for an assessment.

| # Categories | Research validation level |
|---|---|
| All 7 | Gold |
| 5-6 | Silver |
| 3-4 | Bronze |
| 1-2 | Research-based |

7. "Development and validation of instruments to measure learning of expert-like thinking," W. K. Adams and C. E. Wieman, Int. Jour. of Sci. Ed. **33** (9), 1289–1312 (2011). (E)

8. "An introduction to classical test theory as applied to conceptual multiple-choice tests," P. V. Engelhardt, Getting Started in PER **2** (1), 1–40 (2009). (E)

## III. MECHANICS ASSESSMENTS

The topic of mechanics has the largest number of RBAIs, because so many students take introductory mechanics courses at the university level and the content is very standardized. These mechanics RBAIs cover kinematics and forces, energy and rotation (Table III). Because of the wide variety of topics taught in introductory mechanics courses, there is no assessment where all course content is covered. Instead these assessments have a more narrow range of topics, so that you can probe your students' understanding of each sub-topic in mechanics more thoroughly. There is also one mechanics RBA for intermediate and upper-division mechanics courses.

TABLE III. Introductory Mechanics Assessments

| Title | Content | Intended Population | Research Validation | Purpose |
|-------|---------|---------------------|---------------------|---------|
| **Kinematics and Forces** | | | | |
| Force Concept Inventory (FCI) | Kinematics Forces: 1D and 2D | Intro college, high school | Gold | To assess students' understanding of the most basic concepts in Newtonian physics using everyday language and common-sense distractors. |
| Force and Motion Conceptual Evaluation (FMCE) | Kinematics Forces: 1D | Intro college, high school | Gold | To assess students understanding of Newtonian mechanics. |
| Mechanics Baseline Test (MBT) | Kinematics, Forces, Energy, Momentum | Intro college, high school | Bronze | To assess more formal dimensions of basic Newtonian physics. |
| Inventory of Basic Conceptions- Mechanics (IBCM) | Kinematics and Forces | Intro college | Silver | To assess the basic threshold of meaningful understanding of Newtonian theory. |
| Test of Graphical Understanding: Kinematics (TUG-K & TUG-K2) | Kinematics graphs | Intro college, high school (use TUG-K2) | Gold | To assess students' ability to interpret kinematics graphs. |
| Force, Velocity and Acceleration Test (FVA) | Force, Velocity, Acceleration | Intro college | Bronze | To assess students' understanding of the relationships between force, velocity, and acceleration. |
| **Energy** | | | | |
| Energy and Momentum Conceptual Survey (EMCS) | Energy Momentum | Intro college | Gold | To assess conceptual understanding of energy and momentum for standard introductory mechanics courses. |
| Energy Concept Assessment (ECA) | Energy principle, Forms of energy, Work and heat, Absorption & emission spectra, Specification of appropriate systems | Intro college | Silver | To assess conceptual understanding of students in the Matter & Interactions (M&I) Mechanics course. |
| **Rotation** | | | | |
| Rotational and Rolling Motion Conceptual Survey (RRMCS) | Rotational Motion | Intro College | Silver | To assess students' understanding of rotational and rolling motion concepts typically covered in a standard introductory physics course. |
| Rotational Kinematics Inventory (RKI) | Rotational Kinematics | Intro College | Bronze | To assess students' understanding of angular velocity and angular acceleration of a particle in standard introductory physics contexts. |

## A. Kinematics and Forces

### 1. Overview of Kinematics and Forces Assessments

There are six RBAIs which cover kinematics and forces: The Force Concept Inventory[3] (FCI), Force and Motion Conceptual Evaluation[9] (FMCE), Test of Graphical Understanding: Kinematics[10] (TUG-K), Mechanics Baseline Test[11] (MBT), Force, Velocity and Acceleration Test[12] (FVA) and Inventory of Basic Conceptions in Mechanics[13] (IBCM). Research and development of kinematics and forces RBAIs has been occurring since the early 1990s, with the FCI[3] being one of the earliest developed RBAIs in physics. The kinematics and forces RBAIs are all used in introductory classes at the university level, and some are also appropriate for high school students.

The most commonly used test of forces and motion is the Force Concept Inventory[3] (FCI), with results published for over 30,000 students.[1] This is a multiple-choice assessment about the most basic concepts of force and motion. The FCI was the first RBA in physics that presented answer choices consisting of Newtonian concepts and common-sense alternatives that were based on research into student thinking. Physics instructors' first impression of the FCI is often that it is too trivial, but then they are surprised when their students score poorly. Understanding which common-sense alternatives students choose is just as important as looking at the number of correct answers, as this information helps instructors learn how to improve their teaching. The FCI was also the first RBA to show that traditional instruction does not help students learn the most basic concepts in Newtonian physics. The original version of the FCI was a revision of an earlier test called the Mechanics Diagnostic Test[14] (MDT).

There are several variations of the FCI: The Gender FCI[15,16] (or Everyday FCI) uses the same questions and answer choices as the original FCI, but changes the contexts to make them more "everyday" or "feminine." The Animated FCI[17] takes the original FCI questions and animates the diagrams, so it is given on a computer. The Representational Variant of the FCI[18] (R-FCI) takes nine questions from the original

FCI and redesigns them using various representations (such as motion maps, vectorial and graphical representations). The Familiar Context FCI[19] presents the original FCI questions with everyday contexts, e.g., falling fruit instead of stones or colliding shopping carts instead of cars. The Simplified FCI[19] was adapted from the original FCI and made simpler for ninth grade physics.

The Force Motion Conceptual Evaluation[9] (FMCE) is another multiple-choice assessment of forces and motion, that has been widely used (published results for over 10,000 students[1]). The questions on the FMCE are also based on research into student thinking. The FMCE has been used to show that traditional instruction does little to change students' conceptual understanding of forces and motion. Many of the questions on the FMCE have a more complex question format, which includes a description of the problem context, a list of answer choices (often more than 5) and then several questions about that problem situation. In order to give the FMCE in class, a special Scantron with room for 10 answer choices is needed.

Both the FCI and FMCE cover forces and motion, but they have different emphases. The FCI covers more topics than the FMCE, but the FMCE has more questions about each topic to more thoroughly assess students' understanding of each topic. Both tests assess one-dimensional kinematics and Newton's laws. The FCI also includes questions on two-dimensional motion with constant acceleration (parabolic motion), impulsive forces, vector sums, cancellation of forces, and identification of forces.[20] The FMCE includes questions about graphs of motion, whereas the FCI does not. The questions on the FCI each have five answer choices whereas some questions on the FMCE have more than 5. FCI questions 15 and 16 present the same situation as FMCE questions 35-38. FCI question 28 is nearly identical to FMCE question 39. There is a strong correlation between FCI and FMCE scores.[20] Because both of these tests are so popular, there is a large corpus of comparison data, which can help you understand how your students' scores compare to others.

The Mechanics Baseline Test[11] (MBT) assesses more formal dimensions of basic Newtonian physics than the FCI and is meant to be used alongside the FCI to get a well-rounded picture of students' understanding. The answer choices include typical student mistakes but not common-sense alternatives like the FCI. The FCI questions can be answered with no previous physics training, whereas the MBT uses more formal language and includes graphical representations of motion and calculational problems that could not be answered without formal physics training. The MBT covers kinematics and forces, like the FCI, but also includes questions on energy and momentum, which are not covered in the FCI. The MBT includes just a few questions on Newton's first and third laws, since these are well covered in the FCI.

The multiple-choice questions on the Inventory of Basic Conceptions in Mechanics[13] (IBCM) also assesses introductory students' understanding of Newton's laws and forces. The IBCM uses questions from the FCI, MBT, and MDT, but makes slight changes to the wording and answer choices. Since the IBCM takes questions from the FCI and MBT, it is very similar to both of these tests. The IBCM concentrates on Newtonian theory with only two basic models: the free particle and uniformly accelerated motion. It does not include centripetal and centrifugal forces.

The Force, Velocity and Acceleration test[12] (FVA) probes students' understanding of the relationships between force, velocity, and acceleration. Each question presents a scenario with information about either the force, velocity, or acceleration vectors and then asks students about what this means for one of the other vectors. The FVA test provides a coherent picture of student understanding of the relationships between these three by probing six possible conditional relations between them. The relationships between force, velocity, and acceleration on the FVA test are similar to those relationships probed in several questions on the FCI (questions 4, 7, and 9) and FMCE (questions 1, 3, and 12).

The multiple-choice questions on the Test of Graphical Understanding: Kinematics[10] (TUG-K) focus on students' understanding of position, velocity, and acceleration versus time graphs. Questions ask students to find displacement, velocity, or acceleration from a given graph or select a graph corresponding to one given or a textual description. The TUG-K has been validated for high school students, but the TUG-K2 variant was written specifically for high school students. The TUG-K is similar in content to the FMCE, which contains 17 out of 47 questions about graphs of motion, including graphs of force versus time, velocity versus time, and acceleration versus time.

9. "Assessing student learning of Newton's laws: The force and motion conceptual evaluation and the evaluation of active learning laboratory and lecture curricula," R. K. Thornton and D. R. Sokoloff, Am. J. Phys. **66** (4), 338–352 (1998). (E)

10. "Testing student interpretation of kinematics graphs," R. J. Beichner, Am. J. Phys. **62** (8), 750–762 (1994). (E)

11. "A Mechanics Baseline Test," D. Hestenes and M. Wells, Phys. Teach. **30**(3), 159–166 (1992). (E)

12. "Systematic study of student understanding of the relationships between the directions of force, velocity, and acceleration in one dimension," R. Rosenblatt and A. F. Heckler, Phys. Rev. Spec. Top. - Phys. Educ. Res. **7**(2), 1–20 (2011). (I)

13. "Evaluation of the Impact of the New Physics Curriculum on the Conceptual Profiles of Secondary Students," Halloun, I. A. at <http://www.halloun.net/index.php?option=com_content&task=view&id=4&Itemid=6> Beirut, Lebanon, (2007). (E)

14. "The initial knowledge state of college physics students," I. A. Halloun and D. Hestenes, Am. J. Phys. **53**(11), 1043–1055 (1985). (E)

15. "Gender Differences in Student Responses to Physics Conceptual Questions Based on Question Context," L. McCullough, in ASQ Advancing the STEM Agenda in Education, the Workplace and Society, Stout, WI, 1–10 (2011). (E)

16. "Differences in Male/Female Response Patterns on Alternative-format Versions of the Force Concept Inventory," L. McCullough and D. Meltzer, Physics Education Research Conference 2001, Rochester, NY, 103–106 (2001). (E)

17. "Impact of animation on assessment of conceptual understanding in physics," M.

H. Dancy and R. Beichner, Phys. Rev. Spec. Top. - Phys. Educ. Res. **2**(10104), 1–7 (2006). (E)

18.  "Force concept inventory-based multiple-choice test for investigating students' representational consistency," P. Nieminen, A. Savinainen, and J. Viiri, Phys. Rev. Spec. Top. - Phys. Educ. Res. **6**(2), 1–12 (2010). (E)

19.  "Can Assessment of Student Conceptions of Force be Enhanced Through Linguistic Simplification? A Rasch Model Common Person Equating of the FCI and the SFCI," S. E. Osborn Popp and J. C. Jackson, *Annual Meeting of the American Educational Research Association,* San Diego, CA, 1-11 (2009). (I)

20.  "Comparing the force and motion conceptual evaluation and the force concept inventory." R. K. Thornton, D. Kuhl, K. Cummings & J. Marx, Phys. Rev. Spec. Top. - Phys. Educ. Res. **5** (010105), 1–8 (2009). (I)

*2. Recommendations for Choosing a Kinematics and Forces Assessment*

Use the FCI if you want a broad understanding of what your students understand about kinematics and Newton's laws, and lots of comparison data. Use the FMCE if you want a more thorough understanding of what your students understand about kinematics and Newton's laws in one-dimension. Use the MBT in conjunction with the FCI to assess more formal parts of your course. Use the FVA if you want to know about your students' understanding of the relationships between force, velocity, and acceleration vectors. Use the TUG-K if you want to thoroughly assess your students' understanding of motion graphs.

**B. Energy**

*1. Overview of Energy Assessments*

There are two RBAIs that cover energy: the Energy and Momentum Conceptual Survey[21] (EMCS) and the Energy Concept Assessment[22] (ECA). Research and development of energy RBAIs has been occurring since the early 2000s to develop these multiple-choice assessments for introductory classes at the university level.

The Energy and Momentum Conceptual Survey[21] (EMCS) was designed for use in standard first-semester introductory physics courses. It emphasizes energy and momentum in common contexts that your students are likely to have seen in their courses, e.g., carts on tracks, cart filling with rain, bouncing balls, etc.

The Energy Concept Assessment[22] (ECA) was designed specifically to test conceptual understanding of students in the Matter & Interactions (M&I) Mechanics course,[23] which is a first-semester introductory physics course with a radical change in content and emphasis, focusing on the power of fundamental principles, on both the macroscopic and the microscopic level. Because of this, only about half of the questions on the ECA align well with the topics in a standard introductory course. The other half of the questions are not emphasized or covered in a standard course, for example relativistic energy including rest mass, quantized energy levels, and photon emission and absorption.

**21.** "Multiple-choice test of energy and momentum concepts," C. Singh and D. Rosengrant, Am. J. Phys. **71** (6), 607–617 (2003). (E)

**22.** "Designing an Energy Assessment to Evaluate Student Understanding of Energy

Topics," L. Ding, Dissertation (2007). (I)

**23.** "Matter & Interactions," R. Chabay and B. Sherwood, Reviews in PER Vol. 1: Research-based Reform of University Physics, E. F. Redish and P. Cooney, Eds., American Association of Physics Teachers, College Park, MD (2007). (E)

*2. Recommendation for Choosing an Energy Assessment*

The ECA contains questions about non-standard introductory course topics (discussed above) while the EMCS contains more standard questions about energy and momentum. Use whichever test more closely matches the content in your course. Both tests were rigorously developed and tested and found to be reliable.

## C. Rotation

*1. Overview of Rotation Assessments*

There are two tests about rotation: the Rotational and Rolling Motion Conceptual Survey[24] (RRMCS) and the Rotational Kinematics Inventory[25–27] (RKI). Research and development of rotational motion RBAIs has been occurring since the mid-2000s. Both are multiple-choice and are used in introductory classes at the university level.

The Rotational and Rolling Motion Conceptual Survey[24] assesses students' understanding of rotational kinematics and kinetic energy, moment of inertia, torque, and rolling motion. It is appropriate for introductory students in both algebra-based and calculus-based courses.

The Rotational Kinematics Inventory has three parts: 'Part 1: Particles'[25] assesses students' understanding of angular velocity and acceleration of a particle in various standard contexts (the hands of a clock, orbiting plants, swinging pendulum etc.), 'Part 2: Particle in rectilinear motion'[26] assesses students' understanding of the angular velocity and acceleration of a particle moving along a straight line where the origin is not located on that line, 'Part 3: Rigid body about a fixed axis',[27] assesses students' understanding of the rotational kinematics of rigid bodies like pulleys and Ferris wheels. The RKI has been tested with high school students and upper-division college students. Parts of this assessment would also be appropriate for introductory college students. You can use all three parts of the RKI, or only the parts match the content you cover in your course.

The RRMCS and RKI both cover rotational motion topics but with different emphases. The RRMCS focuses on rotational motion concepts commonly taught in introductory courses. The RKI covers these standard topics and also includes some non-standard topics, e.g., a particle in rectilinear motion. Both tests use some physics formalism, which means that the pre-test scores are likely not meaningful because students don't have enough background knowledge to understand the questions. The RRMCS has been used at more institutions and with more students than the RKI.

**24.** "Student understanding of rotational and rolling motion concepts," L. G. Rimoldini and C. Singh, Phys. Rev. Spec. Top. - Phys. Educ. Res. **1**(10102), 1–9 (2005) (E)

25.  "An inventory on rotational kinematics of a particle: unravelling misconceptions and pitfalls in reasoning," K. K. Mashood and V. A. Singh, Eur. J. Phys. **33**(5), 1301–1312 (2012) (E)

26.  "Rotational kinematics of a particle in rectilinear motion: Perceptions and pitfalls," K. K. Mashood and V. A. Singh,  Am. J. Phys. **80**(8), 720–723 (2012) (E)

27.  "Rotational kinematics of a rigid body about a fixed axis: development and analysis of an inventory," K. K. Mashood and V. A. Singh, Eur. J. Phys. **36**(45020), 1–20 (2015) (E)

*2. Recommendations for Rotation Assessments*

Use the RRMCS to assess standard topics in calculus- and algebra-based introductory physics courses and compare to others. Use the RKI if the content matches what you teach in your course.

### D. Intermediate Mechanics

TABLE IV. Intermediate level Mechanics Assessment

| Title | Content | Intended Population | Research Validation | Purpose |
|---|---|---|---|---|
| Colorado Mechanics/Math Methods Instrument (CCMI) | Ordinary differential equations, Taylor series, Potential energy, Simple harmonic motion, Newton's Laws | Intermediate and Upper Division | Silver | To gauge student learning in your first semester classical mechanics course in a way that traditional exams don't allow and compare your students' skills to others. |

The Colorado Mechanics/Math Methods Instrument[28,29] (CCMI) is an open-ended assessment of topics and skills commonly taught in a first-semester intermediate classical mechanics course, including the ability to visualize a problem, correctly apply problem-solving methods, connect math to physics, and describe limiting behavior. The CCMI covers both content and mathematical skills though the questions are largely conceptual, including reasoning, explanation, graphing, and sketching. The CCMI does not cover all content in intermediate classical mechanics, but rather a sample of important skills. CCMI responses are graded using a rubric. This is the only RBA for intermediate classical mechanics.

**28.** "Assessing Student Learning in Middle-Division Classical Mechanics/Math Methods," M. D. Caballero and S. J. Pollock, in Physics Education Research Conference 2013, Portland, OR, 81–84 (2013). (E)

**29.** "Issues and progress in transforming a middle-division classical mechanics/math methods course," S. J. Pollock, R. E. Pepper, & A. D. Marino, AIP Conf. Proc. **1413**, 303–306 (2012). (E)

## IV. ELECTRICITY AND MAGNETISM ASSESSMENTS

TABLE V. Introductory Electricity and Magnetism Assessments

| Title | Content | Intended Population | Research Validation | Purpose |
|-------|---------|---------------------|---------------------|---------|
| **Introductory Electrostatics and Magnetism** | | | | |
| Brief Electricity and Magnetism Assessment (BEMA) | Circuits, Electrostatics, Magnetic Fields and Forces | Intro college | Gold | To assess students' qualitative understanding of basic concepts in electricity and magnetism. |
| Conceptual Survey of Electricity and Magnetism (CSEM) | Electrostatics, Magnetic fields and forces, Faradays Law | Intro college | Silver | To assess students' knowledge about topics in introductory electricity and magnetism. |
| Diagnostic Exam for Introductory Undergraduate Electricity and Magnetism (DEEM) | Electric and magnetic fields and forces, Electrostatic Potential, Maxwell's equations, Induced Currents | Intro college | Bronze | To assess students' understanding of basic concepts of electricity and magnetism. |
| Electricity and Magnetism Conceptual Assessment (EMCA) | Electrostatics, Electric fields and force, Circuits, Magnetism, Induction | Intro college | Bronze | To assess basic concepts in an introductory electromagnetism course, using terms that will feel familiar to students on the pre-test and without overly difficult questions that might discourage students from pursuing physics. |
| Symmetry and Gauss's Law Conceptual Evaluation (SGCE) | Symmetry, electric field, electric flux | Intro College, Upper Division, Graduate | Bronze | To assess students' ability to identify situations where Gauss's Law is applicable and use it to calculate electric field strength. |
| Magnetism Conceptual Survey (MCS) | Magnetic fields and forces, Faraday's Law | Intro college | Silver | To assess difficulties students have with magnetism concepts. |
| **Introductory Circuits** | | | | |
| Determining and Interpreting Resistive Electric Circuits Concepts Test (DIRECT) | DC Circuits | Intro college | Gold | To evaluate students' understanding of direct current (DC) resistive electric circuits concepts. |
| Electric Circuits Conceptual Evaluation (ECCE) | DC and AC circuits | Intro college | Bronze | To assess students' understanding of simple circuit concepts. |
| Inventory of Basic Conceptions- DC Circuits (IBCDC) | DC Circuits | Intro College | Silver | To assess basic conceptions of DC circuits. |

RBAIs on electrostatics and magnetism (E&M) for introductory courses have been around since the late 1990's. There are six research-based assessments that cover electrostatics and magnetism. Four of these are for introductory courses: the Brief Electricity and Magnetism Assessment[30,31] (BEMA), the Conceptual Survey of Electricity and Magnetism[31,32] (CSEM), the Diagnostic Exam for Introductory, Undergraduate Electricity and Magnetism[33] (DEEM), and the Electricity and Magnetism Conceptual Assessment[34] (EMCA). There is one assessment specifically about symmetry and Gauss's law: the Symmetry and Gauss's Law Conceptual Evaluation[35] (SGCE). There is one assessment which covers just magnetism concepts: the Magnetism Conceptual Survey[36] (MCS). There is also the Electromagnetics Concept Inventory (EMCI) suite of assessments which includes EMCI-Waves, EMCI-Fields, and EMCI-Waves and Fields,[37] which were developed for engineering courses and won't be discussed further here.

For circuits, there are three: the Determining and Interpreting Resistive Electric Circuits Concepts Test[38] (DIRECT), the Electric Circuits Conceptual Evaluation[39] (ECCE), and the Inventory of Basic Conceptions- DC Circuits[40] (IBCDC). The CSEM also contains some questions about circuits, but this is not its main focus.

More recently, RBAIs for upper-level courses have been developed. We discuss three: the Colorado Upper Division Electrostatics Diagnostic-Free Response[41] (CUE-FR), the Colorado Upper Division Electrostatics Diagnostics-Coupled Multiple Response[42,43] (CUE-CMR), and the Colorado UppeR-division ElectrodyNamics Test[44] (CURrENT).

Introductory E&M RBAIs are summarized in Table V; upper-level ones are in Table VI.

## A. Introductory Level Electricity and Magnetism

### 1. Electrostatics and magnetism

The two most commonly used RBAIs for introductory electricity and magnetism courses are the Brief Electricity and Magnetism Assessment (BEMA) and the Conceptual Survey of Electricity and Magnetism (CSEM). The BEMA[30] covers the main topics discussed in both the traditional calculus-based E&M physics curriculum and the Matter and Interactions[23] curriculum including basic electrostatics, circuits, magnetic fields and forces, and induction. BEMA questions are mostly qualitative questions with a few semi-quantitative questions, which require simple calculations.

The CSEM[32] is a broad survey of students' knowledge of electricity and magnetism. It aims to assess a range of topics across the standard introductory course content, but without assessing every single topic covered in an introductory course. It is a combination of a test of alternative conceptions and knowledge.  It also has a combination of questions about the phenomena of electricity and magnetism and questions about the formalism explaining the phenomena.

The BEMA and CSEM both cover basic topics covered in introductory electricity and magnetism courses. They share six questions that are identical or nearly identical. CSEM questions have only 5 answer choices, while BEMA questions have up to 10 possible answers choices on some questions. The topics covered on the BEMA and CSEM vary somewhat. The CSEM does not cover circuits, whereas the BEMA does (7 out of 31 questions). CSEM and BEMA scores were compared for one group of students, and on average both pre- and post-test CSEM scores were higher than BEMA scores by 5-6%, a statistically significant difference, with a moderate effect size.[31] But the absolute and normalized gains were similar for the BEMA and CSEM, so for this group of students, both instruments measure learning in a similar way.

There are two other electricity and magnetism tests that haven't been as commonly used and validated: the Electricity and Magnetism Conceptual Assessment (EMCA) and Diagnostic Exam for Introductory, Undergraduate Electricity and Magnetism (DEEM). The multiple-choice conceptual questions on the DEEM[33] measure students' understanding of basic concepts of electricity and magnetism including electric and magnetic fields and force, electrostatic potential and potential energy, Maxwell's equations, and induced currents. The questions align well with the topics commonly taught in an introductory E&M course.

The DEEM is much longer than the CSEM or BEMA (66 questions versus 31 and 32 questions, respectively), so it covers topics much more thoroughly. The DEEM also contains follow-up questions, where students should answer a subsequent question only if they chose a certain answer(s) to a previous question. The DEEM, like the CSEM, does not cover circuits. It also does not cover graphical representations of vector fields, or conductors and insulators. About half the questions on the DEEM ask about the direction of the electric field, magnetic field, velocity, electric potential, or force for different situations.

The EMCA[34] is a multiple-choice assessment of standard second-semester introductory physics concepts including electrostatics, electric fields, circuits, magnetism, and induction. The authors developed the EMCA so that it aligned well with the topics taught in their course and so that it produced similar pre-test scores as the FCI for their student population. The EMCA is easier than the BEMA or CSEM. The authors designed the test this way so that on the pre-test students know the answers to some questions and gain confidence in the course (as opposed to the BEMA and CSEM, which many faculty give only as a post-test because students often score near guessing on the pre-test because they are not familiar with the material), but the post-test can still be used to show mastery at the end of the course.

There is only one assessment specifically about symmetry and Gauss's law: the Symmetry and Gauss's Law Conceptual Evaluation[35] (SGCE) which is designed for students in introductory calculus-based physics, but can also be challenging to upper-level students. The SGCE assesses students' ability to identify situations where Gauss's law is applicable and use it to calculate electric field strength. The SGCE questions are multiple-choice, and primarily conceptual, asking students about when and how to use Gauss's law, but not to explicitly calculate values. The BEMA has one question on Gauss's law, and the CUE-CMR and CUE-FR also ask questions which use Gauss's law and that are aimed at upper-level students.

The Magnetism Conceptual Survey[33] (MCS) was developed to help instructors assess difficulties their students have with magnetism concepts in introductory algebra-based and calculus-based courses. It assesses standard topics in introductory courses up to Faraday's law. The MCS only covers magnetism and not electrostatics, so it follows that it has more questions about magnetism than the BEMA, CSEM, DEEM, or EMCA. The BEMA, CSEM, and MCS all cover charges in magnetic fields and magnetic field from current carrying wires.

**30.** "Evaluating an electricity and magnetism assessment tool: Brief electricity and magnetism assessment," L. Ding, R. Chabay, B. Sherwood, and R. Beichner , Phys. Rev. Spec. Top. - Phys. Educ. Res. **2**(10105), 1–7 (2006). (E)

**31.** "Comparing student learning with multiple research-based conceptual surveys: CSEM and BEMA," S. J. Pollock, AIP Conf. Proc. **1064**, 171–174 (2008). (E)

**32.** "Surveying students' conceptual knowledge of electricity and magnetism," D. P. Maloney, T. L. O'Kuma, C. J. Hieggelke, & A. Van Heuvelen, Am. J. Phys. **69** (7), S12–S23 (2001). (E)

**33.** "Creation of a diagnostic exam for introductory, undergraduate electricity and magnetism," J. D. Marx, Rensselaer Polytechnic Institute, Dissertation (1998). (I)

**34.** "Electricity and magnetism conceptual assessment" M. W. Mccolgan, R. A. Finn, D. Broder & G. Hassel, *in preparation* (2015). (E)

**35.** "Student understanding of symmetry and Gauss's law of electricity," C. Singh, Am. J. Phys. **74**(10), 923–936 (2006). (E)

**36.** "Developing a magnetism conceptual survey and assessing gender differences in student understanding of magnetism," J. Li and C. Singh, AIP Conference Proceedings 1413, 43–46 (2012). (E)

**37.** "Concepty Inventory Assessment Instruments for Electromagnetics Education," B. Notaros, Proceedings of the IEEE Antennas and Propagation Society International Symposium **1**, 684–687 (2002). (E)

*2. Recommendations for Choosing an Electricity and Magnetism Test*

When teaching an introductory electricity and magnetism (E&M) course, use either the CSEM, BEMA, or DEEM. If you would like to assess your students' understanding of circuits in addition to other standard E&M topics, use the BEMA or DEEM. The CSEM and BEMA are used more commonly, so if having comparison data is important to you, use one of these tests. If you want to assess your students' understanding of magnetism separately from other introductory E&M topics, use the MCS. Use the SGCE if you are particularly interested in introductory physics students' understanding of Gauss's law, or if you are making a change to your teaching about Gauss's law and want to understand if that change helped your students.

**B. Circuits**

*1. Overview of Circuits Assessments*

There are three RBAIs of circuits, the Determining and Interpreting Resistive Electric Circuits Concepts Test[38] (DIRECT), the Electric Circuits Conceptual Evaluation (ECCE)[39], and the Inventory of Basic Conceptions-DC Circuits[40] (IBCDC). The CSEM also contains some questions about circuits, but this is not its main focus.

The Determining and Interpreting Resistive Electric Circuits Concepts Test[38] (DIRECT) was developed to evaluate students' understanding of direct current (DC) resistive electric circuits concepts. Overall, the multiple-choice questions weren't designed to assess a particular teaching method, except for a couple of questions about microscopic aspects of circuits, which were closely aligned with the approach used in Electric and Magnetic Interactions (part of the Matter and Interactions curriculum).

The Electric Circuits Concept Evaluation[39] (ECCE) assesses students' understanding of both direct and alternating current circuits. About 80% of the questions are about DC circuits and cover standard concepts around current, voltage, resistance, and brightness of bulbs in circuits containing resistors and capacitors. The remaining 20% of the questions are about AC circuits and ask students to match current versus time graphs to different circuit configurations.[39]

Both the DIRECT and the ECCE ask similar conceptual questions about standard introductory circuits topics, but also ask a couple of questions about non-standard topics: microscopic aspects of current on the DIRECT and AC circuits on the ECCE. Some of the questions on the ECCE have up to 10 answer choices. Also, some of the questions on the ECCE have boxes for students to explain their reasoning. If you grade these short answers, the ECCE could take longer to grade, but many instructors just skip grading these.

The IBCDC is a multiple-choice assessment of DC circuits.[40] The content on the DIRECT and IBCDC is very similar, but the DIRECT has a stronger research base and more comparison data.

**38.** "Students' understanding of direct current resistive electrical circuits," P. V. Engelhardt & R. J. Beichner, Am. J. Phys. **72** (1), 98–115 (2004). (E)

**39.** "Teaching Electric Circuit Concepts Using Microconputer-Based Current/Voltage Problems," D. R. Sokoloff, Microcomputer-Based Labs: Educational Research and Standards, Series F, Computer and Systems Sciences **156**, R. F. Tinker, Ed. 129–146 (1996). (E)

**40.** "Evaluation of the impact of the new physics curriculum on the conceptual profiles of secondary students" I. Halloun, at <http://www.halloun.net/index.php?option=com_content&task=view&id=4&Itemid=6> Beirut, Lebanon, (2007). (E)

### 2. Recommendations for Choosing a Circuits Assessment

Use the DIRECT if you want to assess standard introductory DC circuit concepts because it has a stronger research base and is more commonly used, thus providing you with more comparison data. Use the ECCE if you cover AC circuits and DC circuits. Use the IBCDC if the content matches what you teach in your course more closely.

### C. Upper Level Electricity and Magnetism

**TABLE VI. Upper-Level Electricity and Magnetism Assessments**

| Title | Content | Intended Population | Research Purpose | Validation |
|---|---|---|---|---|
| **Upper-level Electrostatics and Magnetism** | | | | |
| Colorado Upper Division Electrostatics Diagnostic- Free Response (CUE-FR) | Electrostatics, magnetostatics, choosing a problem-solving method | Upper Division | Silver | To assess skills that faculty teaching this course value, such as the ability to visualize a problem, correctly apply problem-solving methods, connect math to physics, and describe limiting behavior, through conceptual questions involving reasoning, explanation, graphing, and sketching. Time intensive to score. |
| Colorado Upper Division Electrostatics Diagnostics- Coupled Multiple Response (CUE-CMR) | Electrostatics, magnetostatics, choosing a problem-solving method | Upper Division | Silver | To assess skills that faculty teaching this course value, such as the ability to visualize a problem, correctly apply problem-solving methods, connect math to physics, and describe limiting behavior, through conceptual questions involving reasoning, explanation, graphing, and sketching. Quick to score. |
| **Upper-level Electrodynamics** | | | | |
| Colorado UppeR-division ElectrodyNamics Test (CURrENT) | Vector calculus, Maxwell's equations, charge and energy conservation, plane waves, transmission and reflection | Upper Division | Bronze | To assess fundamental skills and understanding of core topics from advanced undergraduate electrodynamics. |

*1. Electrostatics and Magnetism*

The Colorado Upper Division Electrostatics Diagnostic-Free Response[41] (CUE-FR) contains open-ended, primarily conceptual questions that assess students' understanding of electrostatics topics (15 out of 17 questions) commonly covered in the first half of a standard upper-division electricity and magnetism course. It also contains two questions about magnetostatics. In addition to assessing E&M content, the CUE-FR assesses several key skills such as the ability to choose a problem-solving method and defend that choice, visualize a problem, connect math to physics, and describe limiting behavior.

The Colorado Upper Division Electrostatics Diagnostic-Coupled Multiple Response[42] (CUE-CMR) was developed to cover the same content as the CUE-FR, but is easier to grade. The questions on the CUE-FR and CUE-CMR are almost identical, but the answer format is different.[43] The CUE-FR has open-ended questions where students show work and explain their reasoning. The CUE-CMR is a coupled multiple-response assessment where students can choose multiple-responses to a given question and are awarded partial credit depending on the accuracy and consistency of their answer. Students are first asked to select the correct answer or easiest method to solve a problem, and then select a 'reasoning element' that supports their initial answer. Students get full credit for selecting all the correct reasoning elements (and only the correct elements). Students can also receive partial credit. A rubric is used to grade the free-responses to the CUE-FR. Partial credit is also granted here. The CUE-FR has 17 questions, while the CUE-CMR has 16 questions (it is missing question 15 from the CUE-FR). On average, students score similarly on the multiple-response version of the test as compared to the free-response version of the test.

41.  "Colorado Upper-Division Electrostatics diagnostic: A conceptual assessment for the junior level," S. V. Chasteen, R. E. Pepper, M. D. Caballero, S. J. Pollock, K. K. Perkins, Phys. Rev. Spec. Top. - Phys. Educ. Res. **8**(20108), 1–15 (2012). (E)

42.  "Multiple-choice Assessment for Upper-division Electricity and Magnetism," B. R. Wilcox & S. J. Pollock, Physics Education Research Conference 2013, Portland, OR, 365–368 (2013). (E)

43.  "Coupled multiple-response versus free-response conceptual assessment: An example from upper-division physics," B. R. Wilcox and S. J. Pollock, Phys. Rev. Spec. Top. - Phys. Educ. Res. **10**(20124), 1–11 (2014) (I)

*2. Recommendations for Choosing an Electricity and Magnetism Test*

If you are teaching an upper-division E&M and want an assessment that is easy to grade and compare to others, use the CUE-CMR. Use the CUE-FR if you want a more in-depth look at the details of your students' reasoning.

*3. Electrodynamics*

There is one assessment of upper-level electrodynamics: the Colorado UppeR-division ElectrodyNamics Test[44] (CURrENT). There is also the Electromagnetics Concept Inventory[37] (EMCI) which includes questions about electrodynamics, but was created for engineering courses, so it will not be discussed further here.

The CURrENT is designed to assess fundamental skills and understanding of core topics in the second semester of junior-level undergraduate electrodynamics covering topics in Chapters 7 through 9 of Griffiths.[45] The CURrENT is free-response in order to assess the ability of upper-level students to generate and justify their own answers. The CURrENT has a conceptual focus, though some mathematical manipulations are required. The CURrENT pre-test contains three questions, while the post-test contains six questions, as students do not have an a-priori familiarity with many of the topics before taking the course. The CURrENT is graded with a rubric. Use the CURrENT to assess your students' understanding in second semester of junior-level undergraduate electrodynamics.

**44.** "Research-based course materials and assessments for upper-division electrodynamics (E&M II)," C. Baily, M. Dubson, and S. J. Pollock, AIP Conf. Proc. **1513**, 54–57 (2013). (E)

**45.** "Introduction to Electrodynamics," D. J. Griffiths, 3rd Edition (Prentice Hall: 2007).

(E)

## V. QUANTUM MECHANICS AND MODERN PHYSICS

There are seven tests covering modern physics and/or quantum mechanics content for sophomore, junior, senior, and graduate level courses. These tests were developed starting in the early 2000's and until very recently. All cover a broad range of topics. These tests are discussed below in groups based on the level of course they are appropriate for. There are two additional graduate quantum mechanics surveys, but these are not research-based and validated, so they will not be discussed further below.[46,47] Intermediate-level tests, such as for Modern Physics courses, are summarized in Table VII; ones for upper-level and graduate courses are in Table VIII.

**46.**   "Graduate Quantum Mechanics Reform," L. D. Carr and S. B. McKagan, Am. J. Phys. **77**(4), 308–319 (2009).(E)

**47.**   "Student understanding of quantum mechanics at the beginning of graduate instruction," C. Singh,  Am. J. Phys. **76**(3), 277–287 (2008). (E)

| Title | Content | Intended Population | Research Validation | Purpose |
|---|---|---|---|---|
| **Relativity** | | | | |
| Relativity Concept Inventory (RCI) | Special Relativity | Intro College | Silver | Measure changes in students' conceptual understanding of special relativity and identify students' misconceptions. |
| **Intermediate Quantum Mechanics** | | | | |
| Quantum Physics Conceptual Survey (QPCS) | Photoelectric effect, Wave particle duality, de Broglie wavelength, Double slit interference, Uncertainty principle | Intro college and Intermediate | Silver | Investigate students' understanding of introductory quantum physics concepts |
| Quantum Mechanics Conceptual Survey (QMCS) | Wave functions, Probability, Infinite square well, One-dimensional tunneling, Wave-particle duality, Energy levels, Uncertainty principle | Intermediate | Silver | Measure the effectiveness of different teaching methods at improving students' conceptual understanding of quantum mechanics, and to use such measurements to improve their teaching. |
| Quantum Mechanics Concept Inventory (QMCI) | Wave functions, Probability, 1D tunneling | Intermediate and Upper-level | Research-based | Assess students' alternative conceptions around 1D potential barriers, tunneling, and probability distributions. |

## A. Modern Physics

### 1. Relativity

The Relativity Concept Inventory[48] is the only RBA that covers special relativity and is for introductory undergraduate courses that cover relevant relativity topics. This is a multiple-choice assessment where students are asked to also rate their confidence for each question. Topics covered include time dilation, length contraction, relativity of simultaneity, inertial reference frames, velocity addition, causality and mass-energy equivalence. Use the RCI if you want to assess your students' conceptual understanding of special relativity and the effectiveness of your instruction.

**48.** "Relativity concept inventory: Development, analysis, and results," J. S. Aslanides and C. M. Savage, Phys. Rev. Spec. Top. - Phys. Educ. Res. **9**(10118), 1–10 (2013).

*2. Intermediate Quantum Mechanics*

There are three tests designed for sophomore-level quantum mechanics: the Quantum Physics Conceptual Survey[49,50] (QPCS), the Quantum Mechanics Conceptual Survey[51] (QMCS) and the Quantum Mechanics Concept Inventory[52] (QMCI). There is one additional quantum assessment, the Quantum Mechanics Visualization Instrument (QMVI), that can be used at multiple levels, including intermediate, upper-level, and graduate quantum, so it will be discussed in the next section. It is not recommended to give any of these tests as a pre-test because they all use basic vocabulary from quantum mechanics that is not meaningful to students until they have some instruction on the topics.

The Quantum Physics Conceptual Survey[49,50] (QPCS) is a conceptual assessment that can be used at the introductory level (if you have covered these topics) and in a sophomore-level modern physics course. There are no equations on the QPCS and most questions focus on wave-particle duality and the photoelectric effect (this is the only quantum test which includes the photoelectric effect). Most of the questions are structured in a way that asks the students about what happens when they do a specific experiment. It was developed in Thailand and tested in Thailand and Australia.

The Quantum Mechanics Conceptual Survey[51] (QMCS) is a highly conceptual multiple-choice assessment for sophomore-level students. It was designed to evaluate the effectiveness of instruction and only covers topics that faculty believe are important in a modern physics course. Some of the questions on the QMCS probe ideas that students have about quantum mechanics, as uncovered in student interviews. For example, one question asks about electrons moving in sinusoidal paths, because interviews found that this is how many undergraduates think about the motion of an electron. Some of the QMCS questions also probe concepts that faculty value. A few of the questions on the QMCS come from other tests (questions 10 and 11 are from the QMVI). Further, the QMCS covers many quantum mechanics topics, but only has 12 questions, so is limited in what it can tell you about what your students learned. The QMCS doesn't explicitly include equations, but it does ask students to think about qualitative relationships in equations.

The Quantum Mechanics Concept Inventory[52] (QMCI) is also very conceptual in nature with no equations included and simple language. Questions are based on students' ideas about quantum as documented in the literature. It is meant for sophomore and junior-level students and was designed to diagnose students' alternative conceptions about quantum mechanics, so each answer choice is associated with a specific alternative conception. Unlike the QMCS, the questions on the QMCI are about a narrow range of topics, with most questions asking about tunneling through one-dimensional barriers. The question format gives statements from a hypothetical student about a given concept and your students have to pick which one they agree with. The QMCI has nine questions, so it is limited in what it tells you about what your students learned.

**49.**   "Probing a deeper understanding of modern physics concepts," T. L. Larkin, P. Meade, and J. Uscinski, 41st ASEE/IEEE 2011 Frontiers in Education Conference, S2H–1–S2H–6 (2011). (E)

**50.**   "Development and Use of a Conceptual Survey in Introductory Quantum Physics," S. Wuttiprom, M. D. Sharma, I. D. Johnston, R. Chitaree, R. & C. Soankwan, International Journal of Science Education **31** (5), 631–654 (2009). (E)

**51.** "Design and validation of the Quantum Mechanics Conceptual Survey," S. B. McKagan, K. K. Perkins, and C. E. Wieman, Phys. Rev. Spec. Top. - Phys. Educ. Res. **6**(20121), 1–17 (2010). (E)

**52.** "Developing a quantum mechanics concept inventory", J. Falk, Uppsala University Dissertation (2004). (I)

*3. Recommendations for Choosing an Intermediate Quantum Mechanics Assessment*

If you are teaching a sophomore-level modern physics course, use the QMCS if you want a broad overview of course topics and the QMCI if you want an in-depth test of one-dimensional potential barriers, tunneling, and probability distribution. Use the QPCS if you want to test photoelectric effect or a more in-depth treatment of wave particle duality. Use QMVI if you want a very detailed look at the relationship between the wave function and shape of potential. The QMVI contains questions from several levels of quantum mechanics, so expect your sophomore-level students to do poorly on most questions.

**B. Upper-level Quantum Mechanics and Beyond**

There are three tests that are designed to assess students' understanding of quantum at the junior level: The Quantum Mechanics Conceptual Assessment[53,54] (QMCA), The Quantum Mechanics Assessment Tool[55] (QMAT), and the Quantum Mechanics Survey[56] (QMS). The Quantum Mechanics Visualization Instrument[57] (QMVI) can be used at several levels and will also be discussed in this section.

The Quantum Mechanics Conceptual Assessment[53,54] (QMCA) is the newest quantum mechanics assessment for a first-semester junior-level quantum mechanics course. It was designed to assess students' understanding of five main topics of quantum measurement: the time-independent Schrödinger equation, wave functions and boundary conditions, time evolution, and probability density. It was also designed to enable comparisons of different teaching methods. The multiple-choice questions on the QMCA were developed using the open-ended questions on the QMAT as a starting point. The QMCA includes math formalism, but most of the questions rely on qualitative understanding of the relationships between equations rather than quantitative calculations. It contains many questions about the Schrödinger equation and a few about measurement as a theoretical construct (e.g. given a wave function, make a measurement, what is the new wave function). There are many questions that use infinite square well potentials and a couple which ask students to think about non-standard potentials qualitatively. The developers recommend using the QMCS as a post-test for sophomore level modern physics classes. It could be used as a pre-test in graduate level quantum to see if students have sufficient conceptual understanding of undergraduate level quantum topics.

The Quantum Mechanics Assessment Tool[55] (QMAT) questions are open-ended and mostly conceptual in nature. It was designed to measure student learning of concepts most valued by faculty, assess students' learning difficulties, and inform course improvement. The content of the QMAT is based on working with faculty to determine learning goals for quantum mechanics. A couple of the questions were taken from an early version of the QMCS. The QMAT questions are a mix of conceptual and math intensive questions, where students are asked to solve equations in some of the questions. There is a rubric for grading the test, but the rubric requires extensive training to get acceptable inter-rater reliability. Further, because this is an open-ended assessment it is difficult to compare results to other institutions. There are limited validation studies of the QMAT, and it has been archived by the developers, so you should use the QMCA, unless you specifically want a short-answer test. Further, the QMCA has been more thoroughly researched and validated.

The Quantum Mechanics Visualization Instrument[57] (QMVI) is a multiple-choice exam and was the first quantum mechanics survey created. It was designed to assess students' understanding of quantum topics at all levels, from sophomore-level to graduate-level. The topics covered are those that authors feel are important for students to learn in the quantum sequence. The QMVI contains 25 questions at all different levels, with very simple questions for sophomore-level students, and very difficult questions for graduate-level students. Because of the variety in difficulty of the questions, it can be used to track students' progress throughout the quantum course sequence. Since it contains questions at the graduate-level, it is a very difficult test. The QMVI contains extensive mathematical formalism. Most of the questions are about the relationship between the shape of the potential and the wave function, with an emphasis on visualizing this relationship. There are a few questions about the uncertainty principle, and two questions about momentum space probability distributions. Some of the questions require 'tricks' to figure out, e.g., making a symmetry argument makes a question very easy, but without the symmetry argument, it is very difficult. The questions are multiple-choice, but also ask students to give a 2-3 line written response and a rating of their confidence level. The developers recommend giving it as an extended take-home exam, as it can take up to two hours for students to complete.

The Quantum Mechanics Survey[56] (QMS) is a multiple-choice assessment for the junior and graduate-level. The QMS was designed to assess students' conceptual understanding of quantum mechanics, but also contains an extensive mathematical formalism. Topics covered on the QMS are those that faculty find important for junior-level quantum mechanics courses. Although students don't have to complete difficult integrals to solve any of the questions, they do need to understand the basics of linear algebra. The QMS has a wide range of topics including wave functions, the expectation value of a physical observable and its time dependence, the role of the Hamiltonian, stationary and non-stationary states and issues related to their time development, and measurements.[56] All questions are restricted to one-dimensional quantum mechanics models.

The content covered by the QMS and QMCA is very similar, but the QMS is more difficult and mathematical than the QMCA, and contains a lot more equations.

53.  "Constructing a Multiple-choice Assessment for Upper-division Quantum Physics from an Open-ended Tool," H. R. Sadaghiani, J. Miller, S. J. Pollock, & D. Rehn, Physics Education Research Conference 2013, Portland, OR, 319–322 (2013). (E)

54.  "Quantum mechanics concept assessment: Development and validation study," H. R. Sadaghiani and S. J. Pollock, Phys. Rev. Spec. Top. - Phys. Educ. Res. **11**(10110), 1–14 (2015). (E)

55.  "Transforming Upper-Division Quantum Mechanics Learning Goals and Assessment," S. Goldhaber, S. J. Pollock, M. Dubson, P. Beale, & K. K. Perkins,

AIP Conf. Proc. **1179** (1), 145–148 (2009). (E)

**56.** G. Zhu and C. Singh, "Surveying students' understanding of quantum mechanics in one spatial dimension," Am. J. Phys. **80**(3), 252–259 (2012). (E)

**57.** "Development and validation of an achievement test in introductory quantum mechanics: the Quantum Mechanics Visualization Instrument (QMVI)," E. Cataloglu, Pennsylvania State University, Dissertation (2002). (E)

*1. Recommendations for Choosing an Upper-Level Quantum Mechanics Assessment*

If you are teaching a junior- or senior-level quantum mechanics course, which test you use depends on both the difficulty level and the range of topics you want to cover: In terms of difficulty, the QMCS is at the lowest level, followed by the QMCI and QPCS, then the QMCA, then the QMVI and QMS. In terms of content, all quantum RBAIs cover some basic ideas about wave functions. The QMVI focuses in great depth on the relationship between the wave function and the shape of the potential. The QPCS is the only assessment that covers the photoelectric effect. The QMCI is entirely conceptual, whereas the QMCA and QMS require some formalism.

**VI. THERMODYNAMICS ASSESSMENTS**

TABLE IX. Thermodynamics Assessments

| Title | Content | Intended Population | Research Validation | Purpose |
|---|---|---|---|---|
| Thermodynamic Conceptual Survey (TCS) | Temperature, heat Transfer, ideal gas law, 1st law of thermodynamics | Intro and Intermediate College | Silver | To assess students' understanding of heat and temperature, the ideal gas law, the first law of thermodynamics and processes. |
| Thermal Concept Evaluation (TCE) | Heat, temperature, heat transfer | Intro college and high School | Silver | To assess introductory college or 3rd-year high school students' understanding and application of thermodynamics concepts using common contexts that reflect students' own conceptions. |
| Heat and Temperature Conceptual Evaluation (HTCE) | Heat, temperature, specific heat capacity, phase changes | Intro College | Bronze | To assess students' understanding of heat and temperature concepts. |

## A. Overview of Thermodynamics Assessments

There are three RBAIs for thermodynamics concepts: The Thermodynamic Conceptual Survey[58] (TCS), Thermal Concept Inventory[59] (TCE), and Heat and Temperature Conceptual Evaluation[60] (HTCE). All of these assessments were developed for introductory level courses. The Thermal and Transport Concept Inventory-Thermodynamics[61] (TTCI-T) and the Thermodynamics Concept Inventory[62] (TCI) were developed specifically for engineering courses, and will not be discussed further here. We are not aware of any research-based assessments on statistical mechanics.

The Thermodynamic Conceptual Survey[58] (TCS) is a multiple-choice conceptual assessment of heat and temperature, the ideal gas law, and the first law of thermodynamics for introductory physics courses. It consists of two parts with part one covering temperature, heat transfer, and the ideal gas law and part two covering the first law of thermodynamics. It is split into two parts so that you can choose the part(s) that most closely match the content covered in your course. The questions on the TCS are all either adapted from other thermodynamics tests or studies of students' understanding of thermodynamics topics. In addition to assessing your students' understanding of these thermodynamics topics, the authors suggest that the questions may be used as teaching materials to help students overcome conceptual difficulties.

The Thermal Concept Inventory[59] (TCE) is a multiple-choice assessment of heat transfer, temperature change, and thermal properties of materials developed based on an inventory of students' alternative conceptions of thermodynamics from the research literature. It was developed for third-year high school students and introductory college students in Australia. The multiple-choice answers allow students to choose from 'everyday physics' answers or 'classroom physics' answers. Many questions consist of a conversation between students and then statements about the opinions of the students involved in the conversation. There are no diagrams or graphs.

The Heat and Temperature Conceptual Evaluation[60,63] (HTCE) is a multiple-choice conceptual assessment of heat, temperature, and heat transfer for introductory physics courses. A majority of the questions are about heat transfer of various materials in cups and about a third have to do with graphing temperature versus time.

The TCS shares many commonalities with the TCE and HTCE because its questions were adapted from various other RBAIs or interview tasks. TCS questions 2, 4, 5, and 6 are the same as TCE questions 8, 11, 14, and 6. TCS questions 1 and 3 are the same as HTCE questions 1 and 8. The TCS covers more thermodynamics concepts than either the TCE or HTCE. TCS includes questions on the first law of thermodynamics and the ideal gas law, whereas neither the TCE nor HTCE contain these topics. Some of the questions on the TCS are also more complex than those on the HTCE or TCE. For example, there is an explanation of a 5-step process of a gas being compressed by a piston, and students are asked questions about work, heat, and energy at various points in the process. The TCS is also the only thermodynamics test that asks students to interpret $P$ vs. $V$ graphs.

The heat and temperature concepts covered on the HTCE are very similar to those covered on the TCE, though the questions on the TCE focus on students' everyday experiences of heat and temperature and many present conversations where students are asked to indicate who they agree with. The HTCE and TCS are more formal and focus on the content of thermodynamics in a physics course. The TCE would be better used as a pre-test, because it focuses on everyday language. The HTCE has three questions about temperature versus time graphs, whereas the TCE has no questions about graphs.

**58.** "Development and Implementation of a Conceptual Survey in Thermodynamics," P. Wattanakasiwich, P. Taleab, M. D. Sharma, & I. D. Johnston, Int. J. Innov. Sci. Math. Educ. **21** (1), 29–53 (2013). (E)

**59.** "Introductory thermal concept evaluation: assessing students' understanding," S. Yeo and M. Zadnik, Phys. Teach. **39**(8), 496–504 (2001). (E)

**60.** "Surveying Thai and Sydney introductory physics students' understandings of heat and temperature," C. Tanahoung, R. Chitaree, C. Soankwan, M. Sharma, & I. Johnston, Proceedings of the Assessment in Science Teaching and Learning Symposium (2006). (E)

**61.** "Rigorous methodology for concept inventory development: Using the "assessment triangle" to develop and test the thermal and transport science concept inventory (TTCI)," R. A. Streveler, R. L. Miller, A. I. Santiago-Román, M. A.

Nelson, M. R. Geist, & B. M. Olds, Int. J. Eng. Educ. **27** (5), 968–984 (2011). (E)

**62.** "Development of Engineering Thermodynamics Concept Inventory instruments," K. C. Midkiff, T. A. Litzinger, and D. L. Evans, in 31st ASEE/IEEE Frontiers in Education Conference, F2A–3, Reno, NV (2001). (E)

**63.** "Surveying Sydney Introduction Physics Students' Understanding of Heat and Temperature," C. Tanahoung, M. D. Sharma, I. D. Johnston, R. Chitaree, & C. Soankwan, Australian Institute of Physics 17th National Congress, Brisbane (2006). (E)

**B. Recommendations for Choosing a Thermodynamics Assessment**

Use the TCS if you want to assess the first law of thermodynamics in addition to other topics such as temperature, heat transfer, phase change, and thermal properties of materials. If you want to use a pre- and post-test use the TCE, because it uses everyday language and ideas that would be familiar to students before a physics course. Further, the format of the TCE where students answer questions about a student discussion could help students get into the frame of mind of discussion and not test taking, which might help you understand their ideas more deeply.

**VII. OPTICS AND WAVES ASSESSMENTS**

TABLE X. Optics and Waves Assessments

| Title | Content | Intended Population | Research Validation | Purpose |
|---|---|---|---|---|
| **Optics** | | | | |
| Four Tier Geometrical Optics Test (FTGOT) | Plane mirrors, Spherical mirrors, Lenses | Intro College | Silver | To assess misconceptions in geometric optics. |
| **Waves** | | | | |
| Mechanical Wave Conceptual Survey (MWCS) | Mechanical waves, wave propagation, wave superposition, wave reflection, standing waves | Intro and Intermediate College, High School | Silver | To identify students' alternative conceptions about mechanical waves before instruction and evaluate the effectiveness of instruction at the end of a course. |
| Wave Diagnostic Test (WDT) | Waves | Intro and Intermediate College, High School | Silver | To understand students' thinking about basic wave concepts. |
| Wave Concept Inventory (WCI) | Visualization of waves, mathematical depiction of waves, and wave definitions | Upper Division | Bronze | To assess students' understanding of wave phenomena in an integrated upper-division engineering course on electronic and electromagnetic topics |

## A. Optics

There is one assessment of geometrical optics, the Four-Tiered Geometrical Optics Test[64] (FTGOT) which is intended for introductory college courses. The FTGOT has `four tiers' of sub-questions for each main question and was developed in Turkey. These ask students to answer a multiple-choice content question, rate their confidence in their answer, indicate their reasoning (also multiple-choice) and then rate their confidence in their reasoning. The test structure can give instructors more confidence that a correct answer to the content question does actually indicate understanding by the student. The questions on the FTGOT ask about observing oneself and observing others with plane mirrors, spherical mirrors, and lenses. Use the FTGOT if you want to assess your students' understanding of geometrical optics concepts at the introductory level.

**64.** "Development and Application of a Four-tiered Test to Assess Pre-service Physics Teachers' Misconceptions about Geometrical Optics," D. Kaltakci, Middle East Technical University, Dissertation (2012).

**B. Introductory Waves Assessments**

There are three RBAIs about waves, two for introductory-level courses, the Mechanical Wave Conceptual Survey[65] (MWCS), the Waves Diagnostic Test[66] (WDT) and one for upper-level courses, the Waves Concept Inventory[67] (WCI).

The Mechanical Wave Conceptual Survey (MWCS)[65] is a multiple-choice assessment of basic wave concepts covered in introductory courses. It has also been tested with high school students. The questions were created based on the open-ended questions from the WDT. The MWCS has four subtopics including propagation, superposition, reflection, and standing waves. Several questions on the MWCS ask students about their reasoning in addition to their answer.

The Waves Diagnostic Test[66] (WDT) has both free-response and multiple-choice questions about mechanical and sound waves topics covered in a typical introductory physics course. The main purpose of the WDT is to learn about students' thinking about waves, not to compare students' scores to a baseline. The WDT elicits rich and varied responses from students that show what they believe about waves and why. This makes the WDT very useful as a benchmark, and allows you to more accurately tailor your instruction to the incoming beliefs of your students. Because the WDT is meant to understand students' thinking, it is not scored. There are two parts to the WDT, and students should complete and turn in part 1 before completing part 2.

The questions on the WDT and MWCS are very similar, since the MWCS was developed from the WDT, but all the questions on the MWCS are multiple-choice, whereas many of the questions on the WDT are free-response. The MWCS is scored in the standard way (% correct) whereas the WDT is meant to be used to understand your students' ideas, and therefore is not scored.

65.  Developing, Evaluating and Demonstrating the Use of a Conceptual Survey in Mechanical Waves. A. Tongchai, M. D. Sharma, I. D. Johnston, K. Arayathanitkul, & C. Soankwan, Int. J. Sci. Educ. **31** (18), 2437–2457 (2009). (E)

66.  "Making Sense of How Students Come to an Understanding of Physics: An Example from Mechanical Waves," M. C. Wittmann, University of Maryland, College Park, Dissertation (1998). (E)

*1. Recommendations for Choosing a Waves Assessment*

Use the MWCS to assess students' understanding of mechanical waves in introductory physics courses if you want to compare students' scores before and after your course with an assessment that is quick and easy to score. Use the WDT for introductory courses if you want to understand your thinking about mechanical waves in a more in-depth way.

## C. Upper-Level Waves Assessments

The Waves Concept Inventory[67] (WCI) is a multiple-choice assessment of upper-division wave phenomenon content including visualization of waves, mathematical depiction of waves, and wave definitions. It was designed to assess the effectiveness of an integrated electrical engineering course covering quantum mechanics and Schrödinger's wave equation as well as Maxwell's wave equations and their application to the propagation of electromagnetic waves, though could also be appropriate for an upper-division physics course. Some of the questions have more than one correct answer, which more thoroughly assess students understanding of the content. There are no calculational questions on the WCI, but students are asked about mathematical equations (e.g. which linear partial differential equation can be used to model wave propagation).

The concepts covered on the WCI are for upper-division engineering courses, though could also be used at the upper-division in a physics department. The WDT and MWCS are meant for introductory courses, so the content and level of these tests are very different. Use the WCI for your upper-division course if the content on the test aligns with what you teach in your class.

**67.** "The wave concepts inventory-an assessment tool for courses in electromagnetic engineering," R. J. Roedel, S. El-Ghazaly, T. R. Rhoads, & E. El-Sharawy, 8th Annu. Front. Educ. Conf. **2**, 647–653 (1998). (E)

## VIII. ASTRONOMY

TABLE XI. Astronomy Assessments

| Title | Content | Intended Population | Research Validation | Purpose |
|---|---|---|---|---|
| **General Astronomy Assessments** | | | | |
| Astronomy Diagnostic Test 2.0 (ADT2) | Seasons, lunar phases, motions in the sky, and size and scale | Intro College | Gold | To assess students' conceptual understanding of introductory astronomy topics. |
| Test of Astronomy Standards (TOAST) | General astronomy content knowledge | Intro College | Silver | To measure students' mastery of core concepts in a general astronomy course. |
| Astronomy Misconceptions Survey (AMS) | Misconceptions about introductory astronomy content | Intro College | Research-based | To identify misconceptions introductory students hold and measure the effectiveness of instruction to dispel these misconceptions. |
| **Specific Astronomy Topic Assessments** | | | | |
| Star Properties Concept Inventory (SPCI) | Stellar properties, Nuclear fusion, Star formation | Intro College | Gold | To measure student learning about the properties and formation of stars. |
| Light and Spectroscopy Concept Inventory (LSCI) | Light, waves, spectroscopy | Intro College | Silver | To measure students' conceptual understanding of topics related to light and spectroscopy, and evaluate the effectiveness of instruction in introductory college astronomy courses. |
| Newtonian Gravity Concept Test (NGCT) | Gravity | Intro College | Silver | To assess student understanding of Newtonian gravity and effectiveness of instruction in general education introductory college astronomy course. |
| Greenhouse Effect Concept Inventory (GECI) | Types of greenhouse gases, energy equilibrium balance, greenhouse effect mechanisms, global warming vs. greenhouse effect | Intro College | Silver | To assess pre- and post-instruction conceptual understanding of the greenhouse effect focusing on the physics of energy flow through Earth's atmosphere. |
| Lunar Phases Concept Inventory (LPCI) | Phases of the moon | Intro College | Bronze | To assess college students' mental models of lunar phases. |

There are eight RBAIs for astronomy, and all are designed for use in the introductory astronomy course. Three of these, the Astronomy Diagnostic Test 2.0[68] (ADT2), the Test of Astronomy Standards[69] (TOAST) and the Astronomical Misconceptions Survey[70] (AMS), contain questions about a wide range of topics covered in an introductory astronomy course and can be used to assess the overall effectiveness of your course. Five of these, the Star Properties Concept Inventory[71] (SPCI), the Light and Spectroscopy Concept Inventory[72] (LSCI), the Newtonian Gravity Concept Inventory[73,74] (NGCI), the Lunar Phases Concept Inventory[75] (LPCI) and the Greenhouse Effect Concept Inventory[76] (GECI) cover a more narrow range of content, and can be used to assess your students' understanding of specific content from your course. All of these RBAIs are multiple-choice. All astronomy assessments are summarized in Table XI.

**A. General Astronomy Assessments**

The Astronomy Diagnostic Test 2.0[68] (ADT2) is a multiple-choice test for non-science majors taking an introductory astronomy course and covers content commonly found in the K-12 curriculum including seasons, lunar phases, motions in the sky, and size and scale. It was designed to help instructors assess their students' initial knowledge coming into a college astronomy course, as the topics included were likely covered in K-12. The ADT2 has been used in many introductory astronomy courses across the US, so there is a lot of comparison data available.

The Test of Astronomy Standards[69] (TOAST) is a multiple-choice broad content assessment of general astronomy content knowledge that is built on and from earlier astronomy assessments (all the astronomy assessments included here). The content on the TOAST was determined based on that which was deemed more important for introductory astronomy students as described in expert position statements from several professional organizations[77,78] and later reviewed by 28 experts in astronomy. This makes it a unique astronomy RBA, as the topics are broad, covering the whole intro course, and are chosen based on based on consensus documents from the astronomy community. Further, most of the questions are taken from other astronomy RBAIs.

The TOAST and ADT2 cover very similar content including phases of the moon, motions in the sky, seasons, scale, distances, sizes, properties and lifecycles of stars, gravity, and the universe. There are several questions that are the same on both tests since the TOAST was created using questions from other astronomy assessments. The TOAST contains questions about production of light (emission, absorption etc.), while both tests ask about the relative speed of electromagnetic waves. There TOAST asks about the Big Bang, and the ADT2 does not. The ADT2 has one question about global warming, and the TOAST does not. Both tests are general assessments for introductory astronomy, and have extensive comparison data available. They have both been well validated.

The Astronomical Misconceptions Survey[70] (AMS) is a multiple-choice survey of common misconceptions in introductory astronomy, e.g, the phases of the moon are caused by the earth's shadow or the seasons are caused by differences in the earth's distance from the sun. There are two versions of the AMS: the true/false version and the multiple-choice version. The true/false version can be used to help instructors understand the misconceptions their students come to their course holding. The multiple-choice version can be given to students to help instructors understand the misconceptions their students have or to assess the effectiveness of different types of instruction at addressing these misconceptions. The questions on the AMS are not about a particular topic, but instead a variety of topics for which students have commonly held incorrect beliefs. Because the AMS is a test of students' misconceptions about astronomy, the topics covered and the focus of the questions is very different from the questions on the ADT2 and TOAST.

**68.** B. Hufnagel, "Development of the Astronomy Diagnostic Test," Astron. Educ. Rev. **1**(1), 47–51 (2002). (E)

**69.** S. J. Slater, "The Development and Validation of The Test of Astronomy STandards (TOAST)," J. Astron. Earth Sci. Educ. **1**(1), 1–22 (2014). (E)

**70.** B. M. C. Lopresto and S. R. Murrell, "An Astronomical Misconceptions Survey," J. Coll. Sci. Teach. **40**(5), 14–22 (2011). (E)

**B. Specific Astronomy Topic Assessments**

The Star Properties Concept Inventory[71] (SPCI) is a multiple-choice assessment of stellar properties, nuclear fusion and star formation. It was developed in response to research on students' alternative conceptions about stars.[79]

The Light and Spectroscopy Concept Inventory[72] (LSCI) is a multiple-choice test about the electromagnetic spectrum and the nature of light meant for introductory astronomy courses. These specific topics have been chosen because they were found to be central topics common across most introductory astronomy courses. The more narrow range of topics means that there are multiple questions probing each. The LSCI questions were developed starting with expert opinions about the important core knowledge around light and the electromagnetic spectrum. Students usually score near guessing (25%) on the pre-test, implying that the LSCI is testing material unfamiliar to students who have not taken an astronomy course. This is different than some other RBAIs where students come in with ideas from everyday life that they use to answer test questions, and then score below the guessing rate. The LSCI has been used in many introductory astronomy courses across the US, so there is a lot of comparison data available.

The Newtonian Gravity Concept Inventory[74] (NGCI) is a multiple-choice assessment of gravity, a foundational topic in introductory astronomy courses. The questions are based on student ideas about gravity and probe four conceptual dimensions including the directionality of gravity, the force law, independence of other forces (e.g. gravity is not affected by rotation) and thresholds related to gravity (e.g. there is not distance for which gravity suddenly stops). The NGCI was developed for use in introductory astronomy courses, but can also be used in introductory physics.

The Lunar Phases Concept Inventory[75] (LPCI) is a multiple-choice assessment of lunar phases concepts including cause and period of lunar phases, period and direction of the Moon's orbit, and observational phenomena. It is designed to assess students' mental models of lunar phases using a mathematical technique called model analysis theory[75]. The result of this analysis is the probability of students in a course answering with the correct model as well as the probability of answering with one of several incorrect models. The LPCI can also be analyzed and scored in the more common way of finding the percent correct on the pre- and post-test and then calculating the normalized gain. Further, since the test content was developed based on students' ideas about the lunar phases, as opposed to expert opinions about the most important content related to lunar phases, it is most appropriate to use the LPCI to understand your students' thinking and mental models, instead of how well their ideas match expert conceptions.

The Greenhouse Effect Concept Inventory[76] (GECI) is a multiple-choice test about the physics of energy flow through Earth's atmosphere. The questions were developed after extensive research on students' beliefs about and models of the greenhouse effect. The GECI is intended to help assess the effectiveness of instruction about the greenhouse effect in introductory astronomy courses.

**71.** "Development of a concept inventory to assess students' understanding and reasoning difficulties about the properties and formation of stars," J. M. Bailey, Astron. Educ. Rev. **6**(2), 133–139 (2008). (E)

72.  "Development and Validation of the Light and Spectroscopy Concept Inventory," E. M. Bardar, E. E. Prather, K. Brecher, & T. F. Slater, Astron. Educ. Rev. **5**(2), 103–113 (2006). (E)

73.  "Development and Calibration of a Concept Inventory to Measure Introductory College Astronomy and Physics Students' Understanding of Newtonian Gravity," K. E. Williamson, Montana State University, Dissertation (2013). (E)

74.  "Development of the Newtonian Gravity Concept Inventory," K. E. Williamson, S. Willoughby, and E. E. Prather,  Astron. Educ. Rev. **12**(1), 010107–1–010107–010120 (2013). (E)

75.  "Developing the Lunar Phases Concept Inventory," R. S. Lindell and J. P. Olsen, Physics Education Research Conference 2002, 1–4, Boise, ID (2002). (E)

76.  "Part I: Development of a Concept Inventory Addressing Students' Beliefs and Reasoning DIfficulties Regarding the Greenhouse Effect Part II: Distribution of Chlorine Measured by Themars Odyssey Gamma Ray Spectrometer," J. M. Keller, University of Arizona, Dissertation (2008). (I)

77.  "National Science Education Standards," National Research Council, Washington DC (1996). (E)

78.  "Project 2061: Benchmarks for Science Literacy," American Association for the Advancement of Science, Washington, DC (1986). (E)

79.  "Development and Validation of the Star Properties Concept Inventory," J. M. Bailey, B. Johnson, E. E. Prather, T. F. Slater, Int. J. Sci. Educ., 1–30 (2011) (E)

## C. Recommendations for Choosing an Astronomy Assessment

Use one of the TOAST or ADT2 if you are making changes to your entire introductory astronomy course, and want to measure the effectiveness of the change. Use the TOAST if you want to assess students' understanding of how light is produced in addition to other standard introductory concepts. Use the ADT2 as a pre-test if you want to understand the ideas your students bring to your course from their K-12 education. Use the AMS if you are particularly interested in understanding your students' misconceptions about astronomy. If instead you are making changes to a specific portion of your course, use an assessment of specific topics that match the content you are changing (SPCI, LSCI, NGCI, LPCI or GECI). The developers of the LSCI point out that the topics covered on the LSCI (electromagnetic spectrum and the nature of light) are foundational and central in many astronomy courses, so you could use this test as a proxy for understanding the effectiveness of your instruction for your course, even though it covers only a subset of the material. Further, if comparing your students' scores to others is important to you, use either the ADT2 or LSCI, as there is a large amount of comparison data published. The TOAST is a newer assessment, so there is less comparison data available now, but this will likely change in the near future.

## XI. CONCLUSION

Table XII summarizes the 46 RBAIs of physics and astronomy content discussed in this resource letter. We have found RBAIs in nearly every major content area in physics, with the exception of statistical mechanics. Most topics have RBAIs at both introductory and the upper levels.

TABLE XII. Summary of RBAs of physics and astronomy content

| Topic | Names | N |
|---|---|---|
| **Mechanics** | | |
| Kinematic and Forces - Intro | FCI, FMCE, MBT, IBCM, TUG-K, FVA | 6 |
| Energy – Intro | EMCS, ECA | 2 |
| Rotation – Intro | RRMCS, RKI | 2 |
| Classical Mechanics - Intermediate | CCMI | 1 |
| **Electricity & Magnetism** | | |
| Electrostatics & magnetism – Intro | BEMA, CSEM, DEEM, EMCA, SGCE, MCS | 6 |
| Circuits – Intro | DIRECT, ECCE, IBCDC | 3 |
| Electricity & magnetism – Intermediate | CUE-FR, CUE-CMR, CURrENT | 3 |
| **Quantum mechanics & modern physics** | | |
| Relativity – Intermediate | RCI | 1 |
| Quantum mechanics – Intermediate | QPCS, QMCS, QMCI | 3 |
| Quantum mechanics – Upper level | QMCA, QMAT, QMVI, QMS | 4 |
| **Thermodynamics** | | |
| Thermodynamics – Intro | TCS, TCE, HTCE | 3 |
| **Optics & Waves** | | |
| Optics – Intro | FTGOT | 1 |
| Waves – Intro | MWCS, WDT | 2 |
| Waves – Upper level | WCI | 1 |
| **Astronomy** | | |
| Astronomy – General – Intro | ADT2, TOAST, AMS | 3 |
| Astronomy – Specific Topics – Intro | SPCI, LSCI, NGCT, GECI, LPCI | 5 |
| **Total** | | **46** |

## ACKNOWLEDGMENTS

We gratefully acknowledge the contributions of the other members of the PhysPort team and KSUPER who worked on this project:  John D. Thompson, Jaime Richards, Devon McCarthy, and Brian Danielak.  This work was partially supported by NSF grants PHYS-1461251, DUE-1256354, DUE-1256354, DUE-1347821, and DUE-1347728.